# Fault Diagnosis for Power Electronics Converters based on Deep Feedforward Network and Wavelet Compression

Lei Kou[a], Chuang Liu[a,1], Guo-wei Cai[a], Zhe Zhang[b]

[a] *School of Electrical Engineering, Northeast Electric Power University, Jilin 132012, China*
[b] *Department of Electrical Engineering, Technical University of Denmark, 2800 Kgs. Lyngby, Denmark*

A B S T R A C T

A fault diagnosis method for power electronics converters based on deep feedforward network and wavelet compression is proposed in this paper. The transient historical data after wavelet compression are used to realize the training of fault diagnosis classifier. Firstly, the correlation analysis of the voltage or current data running in various fault states is performed to remove the redundant features and the sampling point. Secondly, the wavelet transform is used to remove the redundant data of the features, and then the training sample data is greatly compressed. The deep feedforward network is trained by the low frequency component of the features, while the training speed is greatly accelerated. The average accuracy of fault diagnosis classifier can reach over 97%. Finally, the fault diagnosis classifier is tested, and final diagnosis result is determined by multiple-groups transient data, by which the reliability of diagnosis results is improved. The experimental result proves that the classifier has strong generalization ability and can accurately locate the open-circuit faults in IGBTs.

## 1. Introduction

The power electronics converters are widely used in the fields of energy conversion, in which the phase-shifted full-bridge(PSFB) DC-DC converters play a critical role in many cases such as aeronautics, astronautics, hybrid electric vehicles applications and so forth [1]. Power switching tubes is one of the most vulnerable components in power electronic converters because of over-voltage, over-heating or erroneous signal. Although there are various methods to improve the stability of PSFB DC-DC converters, faults are always unavoidable. Meanwhile, the faults of PSFB DC-DC converters may lead to damage of the whole system. Therefore, the reliabilities of these converters are very important [2, 3].

The fault of power semiconductor is mainly shown as short-circuit fault and open-circuit fault [4, 5]. Short-circuit faults, usually protected by standard protection circuit, are considered as the most dangerous, and the IGBT will be shut off immediately once short-circuit fault is detected [6, 7]. On the contrary, though the open-circuit faults are not so destructive, it usually last for some time and may cause secondary damage to other equipments [8-10]. Therefore, the fault diagnosis is of great significance for the stability of the whole system.

Fault diagnosis methods are usually classified into model-based and data-driven methods. Among them, the model-based method is difficult to realize because the fault mathematical model is difficult to establish [11,12]. Compared with model-based methods, machine learning methods are efficient and rely less on the circuit models, which only require the historical data [13,14]. The fault diagnosis method with artificial intelligence (AI)-based techniques for multilevel inverter drive (MLID) was proposed in [15] and [16], in which the neural network fault classifier was trained by the phase voltage after principal component analysis (PCA) dimension reduction, and the classification accuracy of the fault diagnosis classifier can be over 95%. A combined model-based and intelligent method for actuators' small fault detection and isolation was proposed in [17], in which the computational intelligence was used to reduce the influence of uncertainty model and enhance the performance of the diagnostic system. An unsupervised Neural-Network-Based algorithm for three-phase induction motor stator fault was proposed in [18], which requires less accurate mathematical models of the motors. A fault diagnosis algorithm based on multi-state data processing and subsection fluctuation analysis was presented in [19], which can realize multiple open-circuit fault diagnosis for the photovoltaic (PV) inverter. A multiscale adaptive fault diagnosis (MAFD) method based on signal symmetry reconstitution preprocessing (SSRP) was proposed in [20], in which the artificial neural network (ANN) was used to detect the type and location of the switch fault in micro-grid inverter.

Discrete wavelet transform (DWT) can be used to analyze timevarying or non-stationary signals [21]. DWT can automatically extract both low and high frequency components from the signal, while the low frequency component does not contain redundant components [22-23]. A general approach for the transient detection of slip-dependent fault components was proposed in [24], which can identify the wavelet fault currents features. Because the variation of component related to the fault is usually larger than that in steady-state current, the fault diagnosis method with transient current has a potential advantage. A systematic fault diagnosis method for analogue circuits, based on the combination of neural networks and wavelet transforms, was presented in [25], in which the wavelet transforms was used to remove noise from the sampled signals, and PCA was adopted to reduce data dimension.

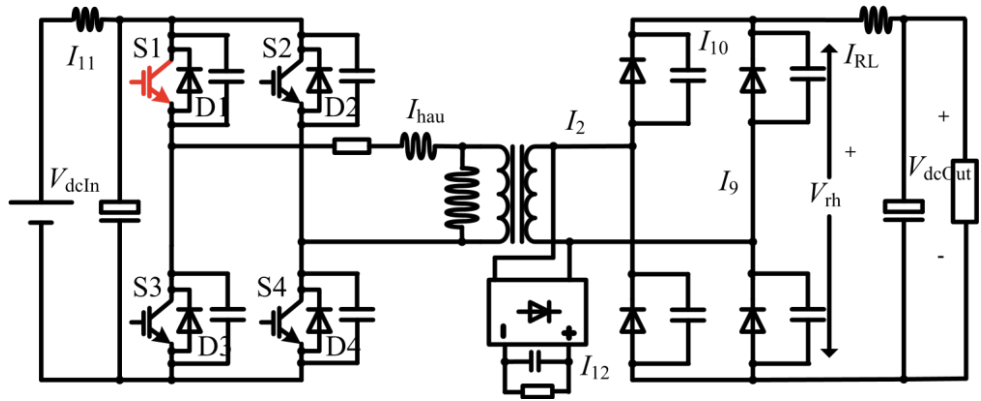

**Fig. 1.** Phase-shifted full-bridge DC-DC converter.

To simplify the study, linear control theory, topology, and ideal semiconductor switch model are still used in power electronic applications. However, the power semiconductor devices are quite different from linearization devices. In actual, it is hard to establish fault mathematical models of many converter topologies. Therefore, it is difficult to apply the mathematical model method to power electronic converter fault diagnosis. With the rapid development of information and computer technology, collecting large amounts of data is no longer a problem, and data-driven methods have been attracted more and more attention [26, 27]. Therefore, a method based on the deep feedforward network (DFN) and wavelet compression algorithm with the transient fault data is proposed in this paper, in which the transient fault features of the power electronic converters are employed to locate the open-circuit faults.

The rest of the paper is organized as follows: Section 2 presents experimental environment and setup, and the transient fault voltage and current data of open-circuit faults in IGBTs of PSFB DC-DC converters are collected from an experimental system. Section 3 describes the proposed fault detection and location method which based on based on DFN and wavelet compression. Section 4 describes the proposed method and experimental results to validate the effectiveness of fault diagnosis method. Conclusions are drawn in the last section.

## 2. Fault data analysis and compression

The proposed method, by which multiple faults can be diagnosed, only requires a small quantity of fault data samples for training fault classifier. The data under various faults are analyzed to find out fewer but more effective features related to fault states.

### 2.1. Fault features selection

The proposed method mainly uses historical data to train deep feedforward network, which has simple network structure and strong approximation capability. The proposed method mainly aims at learning from the data to acquire a strong classification and pattern recognition ability. As shown in Fig. 1, there is an open-circuit fault in IGBT S1 of PSFB DC-DC converter. And the converter has been described in details in [1, 28 and 29].

Therefore, this paper does not describe the topology and control strategy too much.

The sample data are collected form the normal state and the fault states of the PSFB DC-DC converter, and the IGBT faults in S1, S2, S3, S4 are presented as F0, F1, F2, F3 and F4. In order to facilitate computer processing, the states of F0, F1, F2, F3 and F4 are set to 0, 1, 2, 3, 4 as the sample labels, respectively. Fig. 2 shows some original fault data waveform of acquisition features without preprocessing, and the data of each feature in each state are 60,000 samples, respectively. There are 5 states and 300,000 samples for each state. These data will be used for correlation analysis to remove redundant features.

Set a sequence $x_i$ ($i=1,2,…,n$) with $n$ characteristic voltages or characteristic currents and each voltage or current has $m$=300,000 samples, The features $x_i$ ($i=1,2,…,n$) with $n$=8 are given, where $x_1$-$x_8$ represent $I_{hau}$, $I_{RL}$, $I_{12}$, $I_{11}$, $I_9$, $V_{rh}$, $I_{10}$ and $I_2$, respectively. The expression of the covariance between any two sequences is:

$$\mathrm{cov}(x_i, x_j) = \frac{\sum_{k=1}^{m}(x_{ik}-\bar{x}_i)(x_{jk}-\bar{x}_j)}{m-1} \quad (1)$$

$R(i,j)$ is the Pearson correlation coefficient between $x_i$ and $x_j$, and the expression of Pearson correlation coefficient is:

$$R(i,j) = \frac{\mathrm{cov}(x^i, x^j)}{\sigma_{x_i} \sigma_{x_j}} \quad (2)$$

Where $\sigma_{x_i}$ and $\sigma_{x_j}$ are the standard deviations between $x_i$ and $x_j$.

The correlation coefficient between each feature ($x_1$-$x_8$) is calculated, and the matrix is obtained as follows:

$$R = \frac{1}{100}\begin{bmatrix} 100 & 0 & 12 & 100 & 0 & 4 & 62 & 4 & 0 \\ 0 & 100 & 51 & 0 & 100 & 9 & 49 & 0 & \\ & & 64 & 32 & 4 & 9 & 100 & 1 & 79 & 95 \\ & & 0 & 2 & 62 & & 4 & 1 & 100 & 1 & 0 \\ & & 64 & 32 & 4 & 9 & 79 & 1 & 100 & 95 & 68 & 0 & 0 & 0 & 95 & 0 & 95 & 100 \end{bmatrix} \quad (3)$$

As shown in Table 1, both significant correlation and high correlation are considered redundant. According to the correlation coefficient matrix $R$, the feature which has a high correlation with others are removed, and only one of the redundant features is retained. According to this method, on the surface, compared with other features, $x_1$ has a strong classification ability, then $x_1$ removes ($x_5, x_7, x_8$), $x_2$ removes $x_4$, $x_3$ removes $x_6$, then the features ($x_1, x_2, x_3$) are selected. But the experiments result is not good enough because of lack of features. Therefore, fine-tune the features is needed based on the actual situation. Because $x_4$ has the least relevance with other selected features, then the feature $x_4$ is added. Therefore, $I_{11}$, $I_{hau}$, $I_{12}$ and $I_{RL}$ are selected as the features for fault diagnosis. The correlation coefficient matrix of ($x_1, x_2, x_3, x_4$) is as follows:

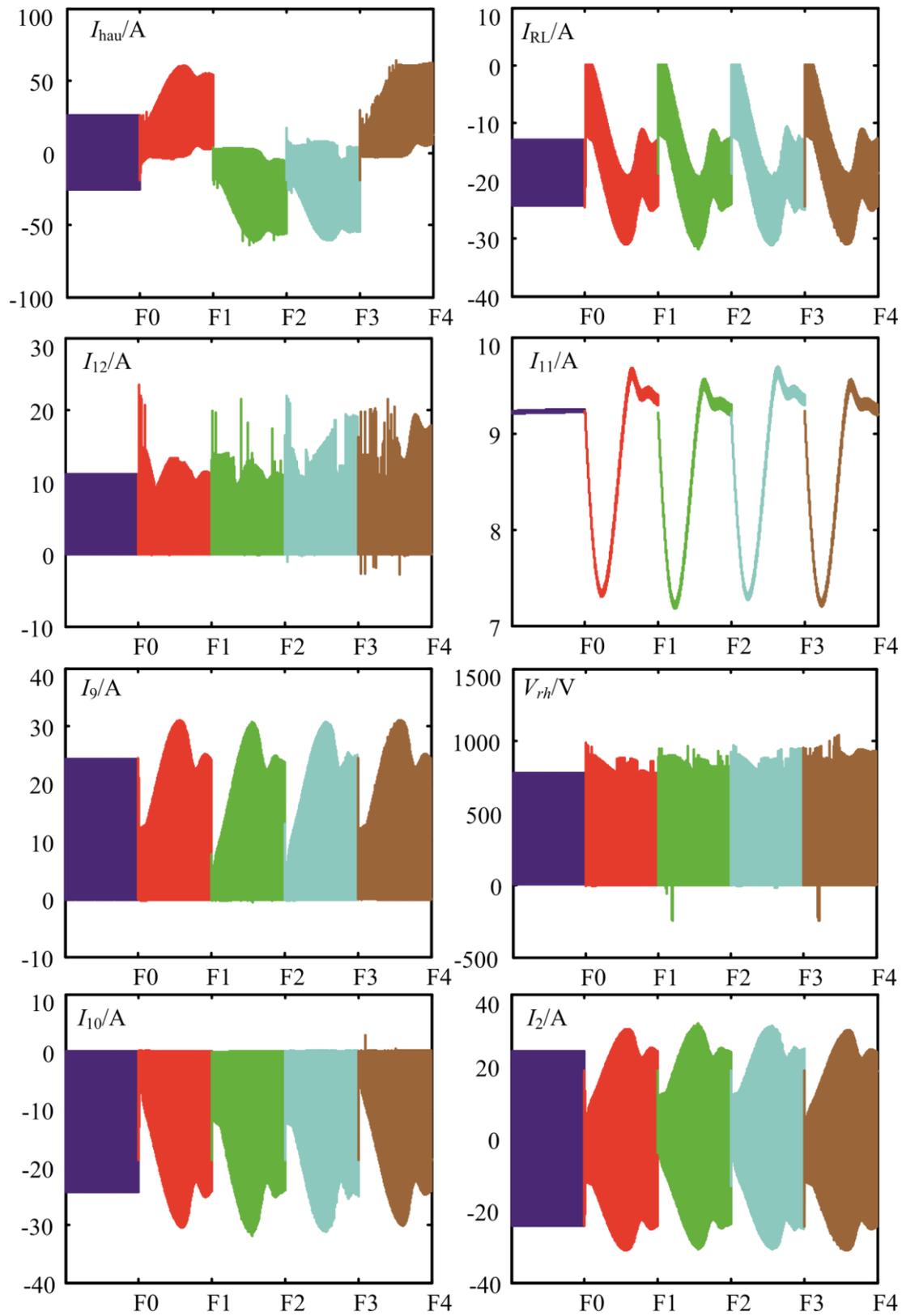

**Fig. 2.** Some original fault data waveform of acquisition features.

**Table 1**
Correlation coefficient and correlation degree.

| Correlation coefficient R | Correlation degree |
|---|---|
| $0 \le |R| < 0.3$ | Weak correlation |
| $0.3 \le |R| < 0.5$ | Moderate Correlation |
| $0.5 \le |R| < 0.8$ | Significant correlation |
| $0.8 \le |R| < 1.0$ | High correlation |

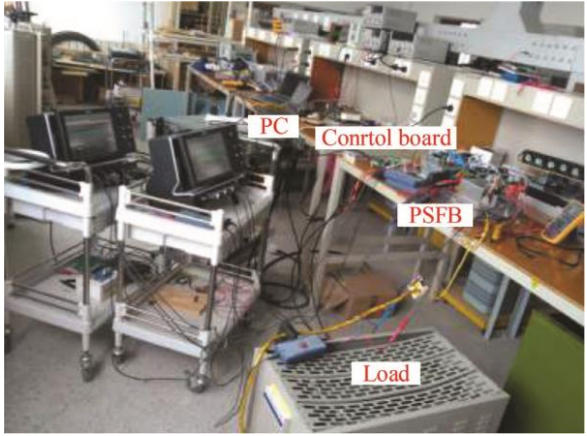

**Fig. 3.** Experimental equipment.

$$R = \begin{bmatrix} 100 & 0 & 0 & 0 \\ 10 & 100 & 12 & 51 \\ 100 & 0 & 12 & 100 & 0 \\ 0 & 51 & 0 & 100 \end{bmatrix} \quad (4)$$

### 2.2. Fault data acquisition

In order to be more consistent with the reality, the fault experiment is directly used to collect sample data. Experimental equipment is shown in Fig. 3. The experimental parameters are: input voltage is 100V, output voltage is 36V, output current is 2.25A, the load is 16Ω, IGBT type is FF300R12KS4, switching frequency 20kHz, filter inductance 490H, the turn ratio of main transformer is 17:17.5, filter capacitance is 20μF, the data sampling frequency of oscilloscope is 125MHz.

Fig. 4 shows the waveform when open-circuit fault occurs in IGBT of PSFB DC-DC converters. As shown in Fig. 4(a)-(d), when open-circuit fault occurs in a single IGBT, the DC output voltage will display a fluctuation but without overvoltage, and the DC output voltage can resume to the normal state within 2ms. PSFB DC-DC converters are easy to run with faults. Therefore, it is important to detect and locate the fault to eliminate the hidden danger timely and effectively.

Fig. 5 shows the waveform of the original data extracted from the state of F0-F4, and each state has 250,000 samples. And the experimental data are exported from the oscilloscope. According to Fig. 2 and Fig. 5, the shape of the experimental data are very close to that of the simulation data.

### 2.3. Data compression based on wavelet compression algorithm

In many applications, low-frequency components are more important than high-frequency components of the signal. The Haar wavelet is easy to implement in hardware, so three levels Haar wavelet is adopted to compress the redundant data in this paper.

Haar wavelet is implemented by wavelet filter banks algorithm: where the length of the input sequence $a^n = \{a_{n,0}, ..., a_{n,2^n-1}\}$ is $2^n$, and the average and detail coefficients of the sequence are obtained. The lowfrequency components $a^{n-1}$ and the high frequency components $d^{n-1}$ ($n$

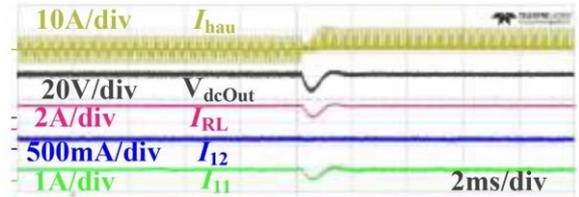
(a) S1 open-circuit fault

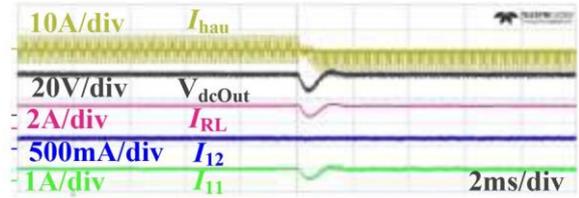
(b) S2 open-circuit fault

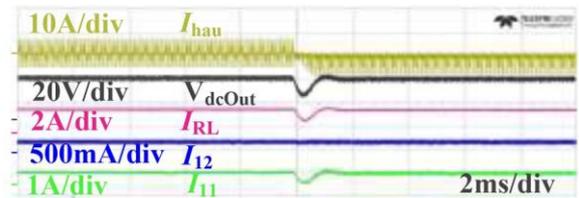
(c) S3 open-circuit fault

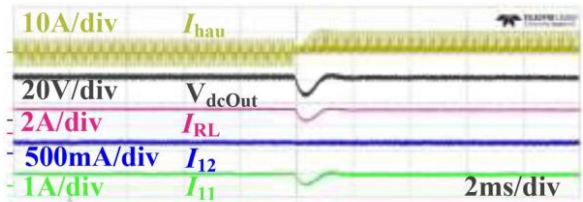
(d) S4 open-circuit fault

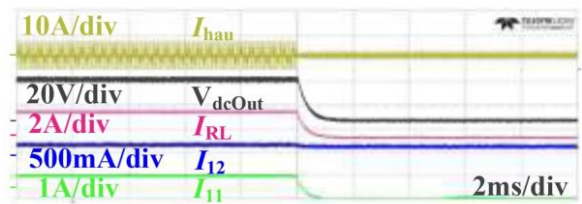
(e) S1 and S2 open-circuit fault

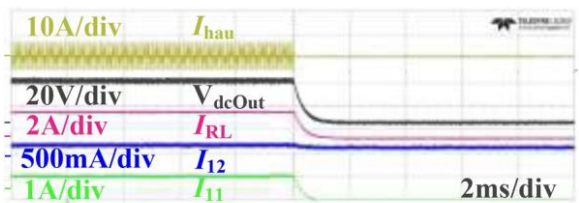
(f) S2 and S4 open-circuit fault

**Fig. 4.** Experimental waveform.

starts from 0) as follows:

$$a^{n-1} = \{a_{n-1,0}, a_{n-1,1}, ..., a_{n-1,2^{n-1}}\}$$

$$d_{n-1} = \{d_{n-1,0}, d_{n-1,1}, ..., d_{n-1,2^n-1}\} \quad (5)$$

Where the expressions of the low-frequency coefficients $a_{n-1,k}$ and the high-frequency coefficients $d_{n-1,k}$ are as follows:

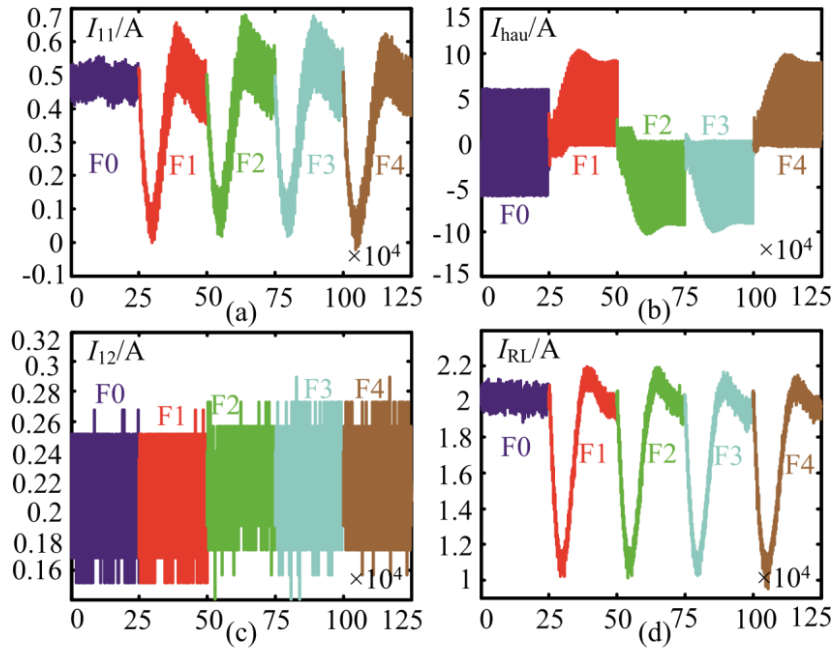

Fig. 5. Actual fault samples under different states.

$$\begin{cases} a_{n-1,k} = (a_{n,k,2} + a_{n,k,2+1})/\sqrt{2}, & k=0, 1, ..., 2^{n-1}-1 \\ d_{n-1,k} = (a_{n,k,2} - a_{n,k,2+1})/\sqrt{2}, & k=0, 1, ..., 2^{n-1}-1 \end{cases} \quad (6)$$

As shown in Table. 2, the averages (low-frequency component) and the detail coefficients (high-frequency component) of the signal are obtained. The high-scale data can be reconstructed from the low-scale average and detail coefficients. Therefore, the original signal can be reconstructed by this method. Similarly, the original data can be compressed from original 300 thousand to 37.5 thousand through 3 levels wavelet transforms.

As shown in Fig. 6, where (a1) and (b1) are the extracted original

**Table. 2**

An example of the one dimensional Haar wavelet transform.

| scale | low-frequency component | high-frequency component |
|---|---|---|
| 3 | [48 34 24 60 72 28 55 121] | |
| 2 | [57.9828 59.3970 70.0036 125.1579] | [9.8995 -25.4558 31.1127 -46.6690] |
| 1 | [83 138] | [-1 -38] |
| 0 | [156.2706] | [-38.8909] |

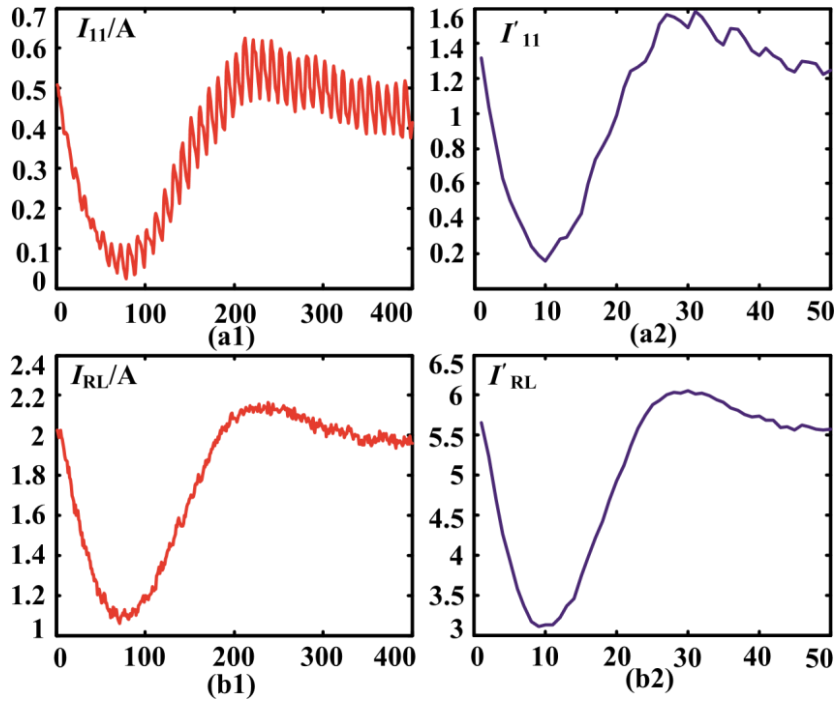

**Fig. 6.** Original fault waveform and waveform after compressed.

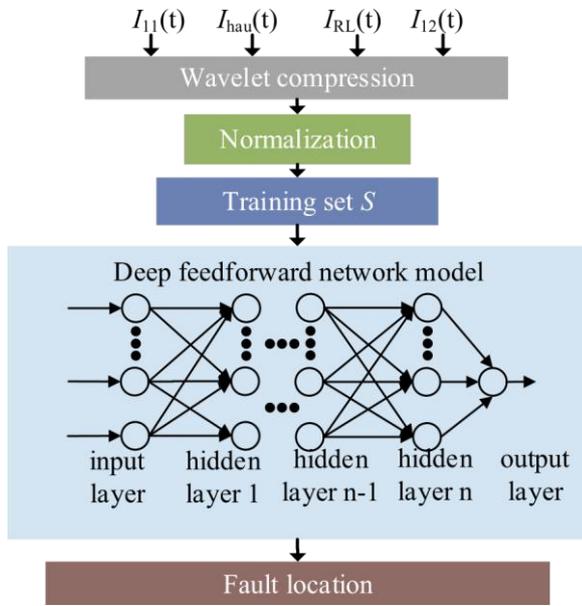

**Fig. 7.** Flow chart of DFN and wavelet compression

**Table 3**
Some samples and diagnostic results.

| Fault location | $I'_{11}$ | $I'_{hau}$ | $I'_{12}$ | $I'_{RL}$ | Actual $f(x)$ | Target $y$ |
|---|---|---|---|---|---|---|
| normal | 1.3255 | −8.8211 | 0.5930 | 5.7867 | 0.0022 | 0 |
| S1 | 1.0152 | 1.1301 | 0.5812 | 5.2623 | 1.0047 | 1 |
| S2 | 0.3368 | −1.4987 | 0.5607 | 3.5319 | 1.9998 | 2 |

**Table 4**
Comparison of fault diagnosis classifiers' training results.

| Method | Sample sizes | Training accuracy | Training time |
|---|---|---|---|
| DFN+original | 1,750,000 | 0.9846 | 12.65h |
| DFN+ wavelet | 218,750 | 0.9850 | 0.52h |
| RF+wavelet | 218,750 | 0.9192 | 0.03h |
| KNN+wavelet | 218,750 | 0.8906 | 0.49h |
| SVM+wavelet | 218,750 | 0.8732 | 0.51h |

waveform, (a2) and (b2) are the waveform after 8 times compression. Although the amount of data is compressed by 8 times, the trend of original data is retained. Therefore, the compressed data can be used to train the fault classifier.

### 3. Training and evaluation of fault diagnosis classifier

DFN algorithm has been widely used in pattern recognition, fault classification and other fields, which is capable of incorporating nonlinearity in the system [30]. The random forest algorithm (RF) was first proposed in [31] by Breiman, which is composed of many decision trees. The RF algorithm has the advantages of train rapidly and can eliminate the issue of over fitting with plenty of features [32]. Compared with DFN and RF algorithm, the fault identification speed of support vector machine (SVM) and K-nearest neighbor (KNN) algorithm are slower, and not conducive to online fault diagnosis [33,34]. This paper makes a comparative analysis of DFN, RF, KNN and SVM algorithm.

**Table 5**
Comparison of fault diagnosis classifiers' test results.

## 3.1. Training of fault diagnosis classifier

According to Fig. 6, the shape of waveform after wavelet compression is similar to that of the original waveform, but the compressed data has no physical meaning because the magnitude of the sample changes. Therefore, the data should be normalized to $[x'_{min}, x'_{max}]$ to reduce the difference in magnitude, where $x'_{min}=-1$ and $x'_{max}=1$, which can reduce the training error and the diagnostic error. The expression of the normalization process is:

$$x' = \begin{cases} \frac{x - \min x}{\max(x_{max} - \min x)(\min x - x_{min})} + x'_{min}, & x_{max} \neq x_{min} \\ x'_{min}, & x_{max} = x_{min} \end{cases} \quad (7)$$

Where $x$ is the sample after wavelet compression, $x_{max}$ and $x_{min}$ are the maximum and minimum value of the characteristic samples in the same group, respectively. And $x'$ is normalized data which can be directly used for training fault classifiers.

The DFN is composed of input layer, hidden layers and output layer. The classification performance of DFN algorithm is associated with the parameters, and the method of setting parameters can be referred to [30]. According to [30], trial and error method, a common method based on the user's experience, could also be used. And the training of fault diagnosis classifier based RF algorithm can be referred to [32]. Fig. 7 shows the flow chart of DFN and wavelet compression, and the parameter settings of the DFN algorithm are as follows: where $[I_{11}, I_{hau}, I_{12}, I_{RL}]$ are the input training samples, the activation function of the hidden layer is *tansig*. The numbers of neurons in each hidden layer are [16,16,16,16,16,16,16,16,16,16]. The activation function of the output layer is *purelin* and the output values are the labels of each state. The generalization training accuracy is set to 0.0001 and the learning efficiency is set to 0.01. The activation function provides a powerful nonlinear modeling capability for DFN algorithm.

The fault diagnosis classifier based on DFN algorithm is trained by $x'$, and then a mature black box pattern classifier $f(x)$ is obtained. The expression is simplified as follows:

$$\begin{cases} round(f(x')), & 0.5 < f(x') < 4.5 \\ y = error\ otherwise \end{cases} \quad (8)$$

$x'$ is the sample obtained after wavelet compression and normalization, which is a combination of $[I'_{11}, I'_{hau}, I'_{12}, I'_{RL}]$, $y$ is the fault diagnosis result of the classifier output after rounding up with *round* function (as shown in Table 3).

## 3.2. Fault diagnosis classifier evaluation

The original training dataset include the data of normal, S1 fault, S2 fault, S3 fault, S4 fault,(S1 and S2) fault and (S2 and S4) fault states, and each state has 250,000 samples, where 30% samples were used for training, and 70% samples for test. The training results and test results are shown in Table. 4 and Table. 5.

According to Table 4, training accuracy of the fault diagnosis classifier based on DFN algorithm with the original data is good, but it needs the most samples and costs the longest training time. The fault diagnosis classifier based on RF and wavelet algorithm takes the shortest time, but its training accuracy is less than DFN and wavelet algorithm. The training time of the KNN and wavelet algorithm, and SVM and wavelet algorithm are about the same as DFN and wavelet algorithm, but their training accuracy are far less than that of DFN and wavelet algorithm. The fault diagnosis classifier based on DFN algorithm and wavelet needs fewer samples, shorter training time and the highest training accuracy among them.

The identification precision rate and recall rate of different fault diagnosis classifiers are listed in Table. 5. According to Table. 5, the fault diagnosis classifier based on DFN and wavelet algorithm gets the best performance, and that of SVM and wavelet algorithm is the worst. Based on the comparison results and the description of RF algorithm in [30], it is confirmed that the classifier based on DFN algorithm performs better than the classifier based on RF algorithm under the condition of fewer types of features. However, the fewer types of features needed the better. Therefore, the classifier based on DFN algorithm and wavelet compression is adopted to fault diagnosis for PSFB DC-DC converters.

## 4. Fault diagnosis experiments for PSFB DC-DC converters

The fault diagnosis experiments are implemented to verify the effectiveness of fault diagnosis classifier which is based on DFN algorithm and wavelet compression. The experimental platform (as shown in Fig. 3) mainly includes a PSFB DC-DC converter, an ARM controller (STM32F103), a FPGA controller (EP4CE115F23), an A/D converter chip (ADS1256), and a PC which can be substituted with a low-cost industrial computer. The trained fault diagnosis classifier which runs on the PC is employed to monitor and diagnose the state of PSFB DC-DC converters, and the fault diagnosis process is shown in Fig. 7. The opencircuit fault experiment is simulated by turning off the control signal of IGBT in PSFB DC-DC converters. As shown in Fig. 8, the experiments are carried out with S1 and S3 open-circuit faults as examples.

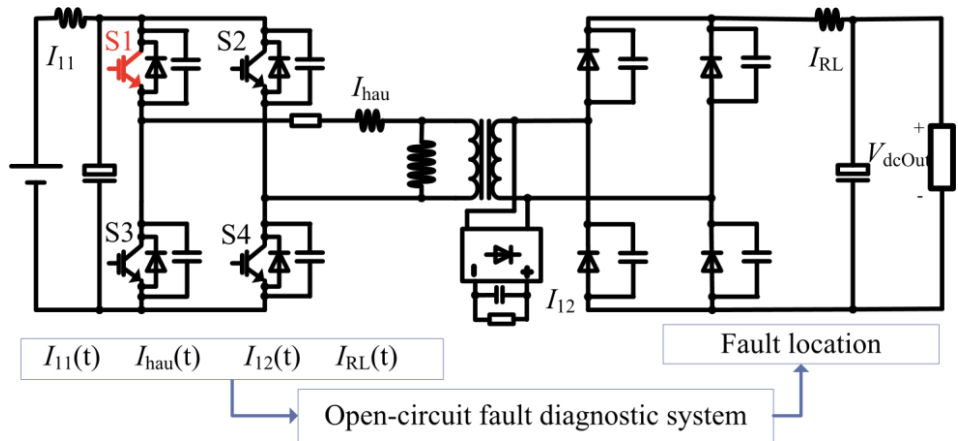

**Fig. 8.** Fault Diagnosis System for PSFB DC-DC converter

| Fault location | DFN+original | | DFN+ wavelet | | RF+ wavelet | | KNN+wavelet | | SVM+wavelet | |
|---|---|---|---|---|---|---|---|---|---|---|
| | precision | recall | precision | recall | precision | recall | precision | recall | precision | recall |
| normal | 0.9835 | 0.9989 | 0.9441 | 0.9935 | 0.7633 | 0.9673 | 0.7601 | 0.9543 | 0.7313 | 0.9503 |
| S1 | 0.9187 | 0.9827 | 0.8618 | 0.9843 | 0.7740 | 0.9495 | 0.6978 | 0.9201 | 0.6253 | 0.9162 |
| S2 | 0.7500 | 0.9740 | 0.8210 | 0.9742 | 0.2738 | 0.8488 | 0.2623 | 0.8549 | 0.2379 | 0.8473 |
| S3 | 0.8184 | 0.9731 | 0.8592 | 0.9701 | 0.5159 | 0.8735 | 0.5163 | 0.8652 | 0.4520 | 0.8348 |
| S4 | 0.8841 | 0.9748 | 0.8818 | 0.9847 | 0.6956 | 0.9570 | 0.6400 | 0.9302 | 0.6319 | 0.8756 |
| S1 and S2 | 0.9547 | 0.9730 | 0.9844 | 0.9820 | 0.9930 | 0.8488 | 0.9151 | 0.8472 | 0.8910 | 0.8341 |
| S2 and S4 | 0.9307 | 0.9750 | 0.9652 | 0.9827 | 0.9818 | 0.8735 | 0.9182 | 0.8625 | 0.9064 | 0.8543 |

The sampling frequency of the control system is set to 16kHz, it only needs to send 20 points per 10ms to the fault diagnosis classifier after Haar wavelet compression on the controller, and it can reduce the pressure on data transmission over the network. Fig. 9 shows the waveform after wavelet compression, each time 20 sets fault samples are used for fault diagnosis, and 20 diagnostic results are output. And the corresponding diagnostic results are shown in Fig. 10.

Fig. 10 shows the 20 diagnostic results every time and the 20 results jointly decide the fault location. Fig. 10 (a) shows the diagnostic results when the open-circuit fault occurs in IGBT S1, where the samples from 1 to 14 are in normal state, and from 15 to 20 are S1 fault state, and then the fault location is S1. Fig. 10 (b) shows the diagnostic results when the open-circuit fault occurs in IGBT S3, where the samples from 1 to 15 are in normal state, and from 16 to 20 are S3 fault state, and then the fault location is S3.

Therefore, based on the results shown in Fig. 10, it is confirmed that the fault diagnosis classifier based on DFN algorithm and wavelet compression can be adopted to locate the fault IGBT in PSFB DC-DC converters.

## 5. Conclusion

This paper presented a fault diagnosis method for power electronics converters based on DFN and wavelet compression with transient fault samples. The magnitude change of fault related components in transient state is usually larger than that in steady state. Therefore, transient features are used to train fault diagnosis classifier in this paper.

The redundant features are reduced by the way of correlation analysis, and then the additional circuit is also reduced as much as possible. At the same time, the Haar wavelet is easy to implement in hardware, and it can be used to compress samples and filter signals. In this way, the redundant data and noise are removed, the training speed of DFN is greatly accelerated, and the fault diagnosis accuracy has also been improved.

Finally, the fault diagnosis experiments have been carried out to verify the effectiveness of the proposed method. It reduces the dependence on the fault mathematical model of the power electronics converters. The fault diagnosis classifier based DFN and wavelet compression has the advantages of good classification ability, high computational efficiency and implementation simplicity. The proposed method can locate the open-circuit fault in IGBT accurately.

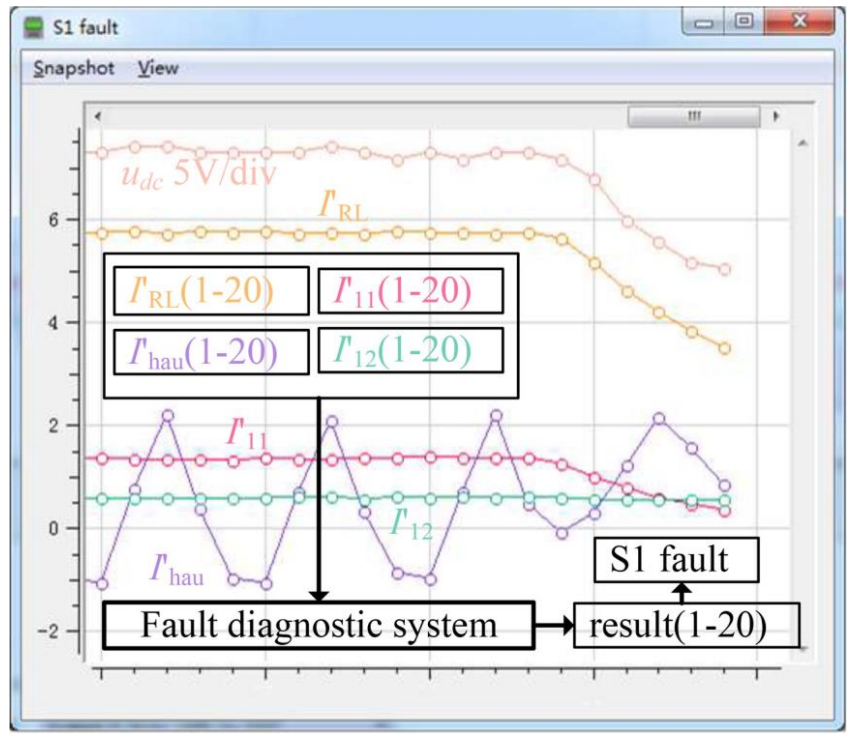

(a) S1 open-circuit fault

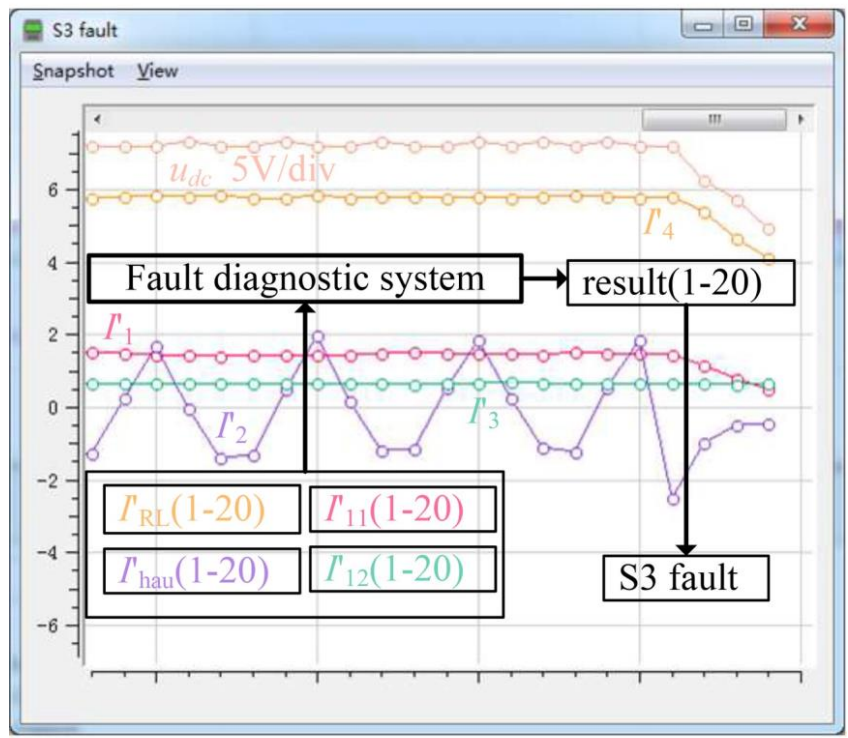

(b) S3 open-circuit fault

**Fig. 9.** Fault diagnosis Experiments

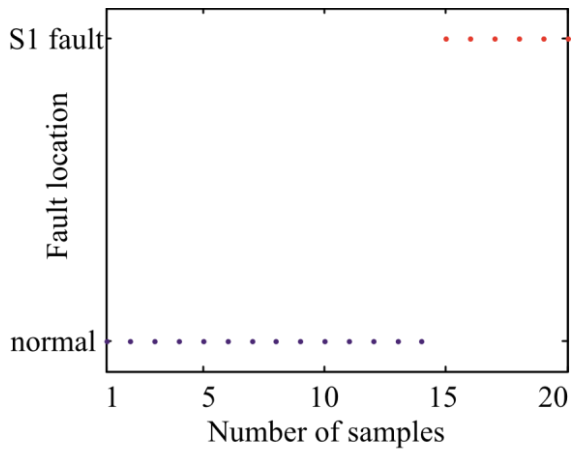

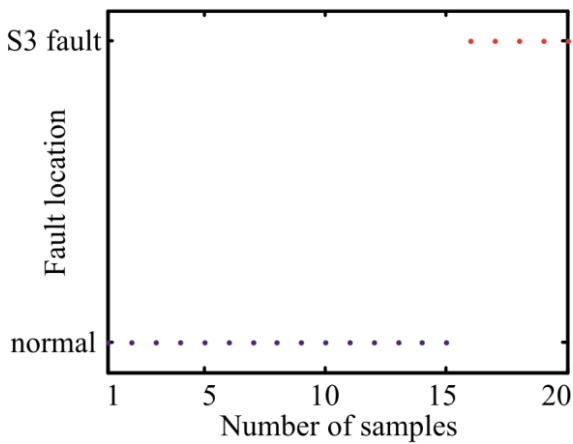

**Fig. 10.** Fault diagnosis results

**Acknowledgements**

This research is funded by National Key R&D Program of China under grant number 2017YFB0903300.

**References**

[1] Pei X, S Nie, Y Chen, et al., Open-Circuit Fault Diagnosis and Fault-Tolerant Strategies for Full-Bridge DC–DC Converters, IEEE Transactions on Power Electronics 27 (5) (2012) 2550–2565.
[2] J. Wang, H. Ma, Z. Bai, A Submodule Fault Ride-Through Strategy for Modular Multilevel Converters With Nearest Level Modulation, IEEE Transactions on Power Electronics, 33 Feb. 2018, pp. 1597–1608.
[3] B. Li, S. Shi, B. Wang, G. Wang, W. Wang, D. Xu, Fault Diagnosis and Tolerant Control of Single IGBT Open-Circuit Failure in Modular Multilevel Converters, IEEE Transactions on Power Electronics, 31 April 2016, pp. 3165–3176.
[4] A. Sai Sarathi Vasan, B. Long, M. Pecht, Diagnostics and Prognostics Method for Analog Electronic Circuits, IEEE Transactions on Industrial Electronics, 60 Nov. 2013, pp. 5277–5291.
[5] M. Aly, E.M. Ahmed, M. Shoyama, A New Single-Phase Five-Level Inverter Topology for Single and Multiple Switches Fault Tolerance, IEEE Transactions on Power Electronics, 33 Nov. 2018, pp. 9198–9208.
[6] M. Salehifar, R. Salehi Arashloo, M. Moreno-Eguilaz, V. Sala, L. Romeral, Observerbased open transistor fault diagnosis and fault-tolerant control of five-phase permanent magnet motor drive for application in electric vehicles, IET Power Electronics, 8 1 2015, pp. 76–87.
[7] U. Choi, H. Jeong, K. Lee, F. Blaabjerg, Method for Detecting an Open-Switch Fault in a Grid-Connected NPC Inverter System, IEEE Transactions on Power Electronics, 27 June 2012, pp. 2726–2739.
[8] I. Jlassi, J.O. Estima, S.K. El Khil, N.M. Bellaaj, A.J.M. Cardoso, A Robust Observer-Based Method for IGBTs and Current Sensors Fault Diagnosis in Voltage-Source Inverters of PMSM Drives, IEEE Transactions on Industry Applications, 53 MayJune 2017, pp. 2894–2905.
[9] M. Chen, D. Xu, X. Zhang, N. Zhu, J. Wu, K. Rajashekara, An Improved IGBT ShortCircuit Protection Method With Self-Adaptive Blanking Circuit Based on $V_{CE}$ Measurement, IEEE Transactions on Power Electronics, 33 July 2018, pp. 6126–6136.
[10] S. Shao, A.J. Watson, J.C. Clare, P.W. Wheeler, Robustness Analysis and Experimental Validation of a Fault Detection and Isolation Method for the Modular Multilevel Converter, IEEE Transactions on Power Electronics, 31 May 2016, pp. 3794–3805.
[11] Z. Gao, C. Cecati, S.X. Ding, A Survey of Fault Diagnosis and Fault-Tolerant Techniques—Part I: Fault Diagnosis With Model-Based and Signal-Based Approaches, IEEE Transactions on Industrial Electronics, 62 June 2015, pp. 3757–3767.
[12] Y. Chen, Z. Li, S. Zhao, X. Wei, Y. Kang, Design and Implementation of a Modular Multilevel Converter With Hierarchical Redundancy Ability for Electric Ship MVDC System, IEEE Journal of Emerging and Selected Topics in Power Electronics, 5 March 2017, pp. 189–202.
[13] M. Aly, E.M. Ahmed, M. Shoyama, A New Single-Phase Five-Level Inverter Topology for Single and Multiple Switches Fault Tolerance, IEEE Transactions on Power Electronics, 33 Nov. 2018, pp. 9198–9208.
[14] J. Poon, P. Jain, I.C. Konstantakopoulos, C. Spanos, S.K. Panda, S.R. Sanders, Model-Based Fault Detection and Identification for Switching Power Converters, IEEE Transactions on Power Electronics, 32 Feb. 2017, pp. 1419–1430.
[15] S. Khomfoi, L.M. Tolbert, Fault Diagnosis and Reconfiguration for Multilevel Inverter Drive Using AI-Based Techniques, IEEE Transactions on Industrial Electronics, 54 Dec. 2007, pp. 2954–2968.
[16] S. Khomfoi, L.M. Tolbertt, B. Ozpineci, Cascaded H-bridge Multilevel Inverter Drives Operating under Faulty Condition with AI-Based Fault Diagnosis and Reconfiguration, 2007 IEEE International Electric Machines & Drives Conference, Antalya, 2007, pp. 1649–1656.
[17] Y. Cheng, R. Wang, M. Xu, A Combined Model-Based and Intelligent Method for Small Fault Detection and Isolation of Actuators, IEEE Transactions on Industrial Electronics, 63 April 2016, pp. 2403–2413.
[18] J.F. Martins, V. Ferno Pires, A.J. Pires, Unsupervised Neural-Network-Based Algorithm for an On-Line Diagnosis of Three-Phase Induction Motor Stator Fault, IEEE Transactions on Industrial Electronics, 54 Feb. 2007, pp. 259–264.
[19] Z. Huang, Z. Wang, H. Zhang, Multiple Open-Circuit Fault Diagnosis Based on Multistate Data Processing and Subsection Fluctuation Analysis for Photovoltaic Inverter, IEEE Transactions on Instrumentation and Measurement, 67 March 2018, pp. 516–526.
[20] Z. Wang, Z. Huang, C. Song, H. Zhang, Multiscale Adaptive Fault Diagnosis Based on Signal Symmetry Reconstitution Preprocessing for Microgrid Inverter Under Changing Load Condition, IEEE Transactions on Smart Grid, 9 March 2018, pp. 797–806.
[21] X. Chen, Real Wavelet Transform-Based Phase Information Extraction Method: Theory and Demonstrations, IEEE Transactions on Industrial Electronics, 56 March 2009, pp. 891–899.
[22] A. Bouzida, O. Touhami, R. Ibtiouen, A. Belouchrani, M. Fadel, A. Rezzoug, Fault Diagnosis in Industrial Induction Machines Through Discrete Wavelet Transform, IEEE Transactions on Industrial Electronics, 58 Sept. 2011, pp. 4385–4395.
[23] J. Pons-Llinares, J.A. Antonino-Daviu, M. Riera-Guasp, M. Pineda-Sanchez, V. Climente-Alarcon, Induction Motor Diagnosis Based on a Transient Current Analytic Wavelet Transform via Frequency B-Splines, IEEE Transactions on Industrial Electronics, 58 May 2011, pp. 1530–1544.
[24] M. Riera-Guasp, J.A. Antonino-Daviu, M. Pineda-Sanchez, R. Puche-Panadero, J. Perez-Cruz, A General Approach for the Transient Detection of Slip-Dependent Fault Components Based on the Discrete Wavelet Transform, IEEE Transactions on Industrial Electronics, 55 Dec. 2008, pp. 4167–4180.
[25] Y. He, Y. Tan, Y. Sun, Wavelet neural network approach for fault diagnosis of analogue circuits, IEE Proceedings - Circuits, Devices and Systems, 151 12 Aug. 2004, pp. 379–384.
[26] S. Yin, S.X. Ding, X. Xie, H. Luo, A Review on Basic Data-Driven Approaches for Industrial Process Monitoring, IEEE Transactions on Industrial Electronics, 61 Nov. 2014, pp. 6418–6428.
[27] L. Wen, X. Li, L. Gao, Y. Zhang, A New Convolutional Neural Network-Based DataDriven Fault Diagnosis Method, IEEE Transactions on Industrial Electronics, 65 July 2018, pp. 5990–5998.
[28] L. Zhu, A Novel Soft-Commutating Isolated Boost Full-Bridge ZVS-PWM DC–DC Converter for Bidirectional High Power Applications, IEEE Transactions on Power Electronics, 21 March 2006, pp. 422–429.
[29] Z. Wang, Y. Wang, Y. Rong, Design of closed-loop control system for a bidirectional full bridge DC/DC converter, Applied Energy 194 (15) (May. 2016) 617–625.
[30] F Zhao, Y Zhang, H Chen, Multimodal Data Analysis and Integration for Multi-Slot Spectrum Auction Based on Deep Feedforward Network[J], Pattern Recognition (2017).
[31] L. Breiman, Random Forests[J], Machine Learning 45 (1) (2001) 5–32.
[32] AM Shah, BR Bhalja, Fault discrimination scheme for power transformer using random forest technique[J], IET Generation, Transmission & Distribution (2016).
[33] L. Cai, N.F. Thornhill, S. Kuenzel, B.C. Pal, Wide-Area Monitoring of Power Systems Using Principal Component Analysis and k-Nearest Neighbor Analysis, IEEE Transactions on Power Systems 33 (5) (Sept. 2018) 4913–4923.

[34] SK Jain, SN Singh, Harmonics estimation in emerging power system: Key issues and challenges[J], Electric Power Systems Research 81 (9) (2011) 1754–1766.